\documentclass[structabstract]{aa}
\usepackage{graphicx}
\usepackage{txfonts}
\usepackage{appendix}
\usepackage{natbib}
\usepackage{longtable, portland}
\usepackage{supertabular}
\usepackage{rotating}
\bibpunct{(}{)}{;}{a}{}{,}
\begin{document}
   \title{Evolution of the Fundamental Plane of 0.2$<$z$<$1.2 Early--type galaxies in the EGS}

   \author{M. Fern\'andez Lorenzo\inst{1}$^,$\inst{2}, J. Cepa\inst{1}$^,$\inst{2}, A. Bongiovanni\inst{1}$^,$\inst{2}, A.M. P\'erez Garc\'ia\inst{1}$^,$\inst{2}, A. Ederoclite\inst{1}$^,$\inst{2}, M.A. Lara-L\'opez\inst{1}$^,$\inst{2}, M. Povi\'c\inst{1}$^,$\inst{2}, \and  M. S\'anchez-Portal\inst{3}
          }

   \institute{Instituto de Astrof\'isica de Canarias (IAC),
              C/ V\'ia L\'actea S/N, 38200 La Laguna, Spain\
   \and
     Departamento de Astrof\'isica, Universidad de La Laguna, Spain
         \and
             Herschel Science Centre, INSA/ESAC, Madrid, Spain
            }

 \date{Received.....; accepted..... }


  \abstract
   {The Fundamental Plane relates the structural properties of early--type galaxies such as its surface brightness and effective radius with its dynamics. The study of its evolution has therefore important implications for models of galaxy formation and evolution.}
   {This work aims to identify signs of evolution of early--type galaxies through the study of parameter correlations such as the Fundamental Plane, the Kormendy, and the Faber--Jackson relations, using a sample of 135 field galaxies extracted from the Extended Groth Strip in the redshift range 0.2$<$z$<$1.2.}
   {Using DEEP2 data, we calculate the internal velocity dispersions by extracting the stellar kinematics from absorption line spectra, using a maximum penalized likelihood approach. Morphology was determined through visual classification using the V+I images of ACS. The structural parameters of these galaxies were obtained by fitting de Vaucouleurs stellar profiles to the ACS I--band images, using the ${\tt GALFIT}$ code. To check the effect on the Fundamental Plane of the structural parameters, S\'ersic and bulge--to--disc decomposition models were fitted to our sample of galaxies. A good agreement was found in the Fundamental Plane derived from the three models.}
   {Assuming that effective radii and velocity dispersions do not evolve with redshift, we have found a brightening of 0.68 mag in the B--band and 0.52 mag in the g--band at $<$z$>$=0.7. However, the scatter in the FP for our high--redshift sample is reduced by half when we allow the FP slope to evolve, suggesting a different evolution of early--type galaxies according to their intrinsic properties, such as total mass, size or luminosity. The study of the Kormendy relation shows the existence of a population of very compact (R$_e$$<$2 Kpc) and bright galaxies ($-$21.5$>$M$_g$$>$$-$22.5), of which there are only a small fraction (0.4$\%$) at z = 0. Studying the luminosity--size and stellar mass--size relations, we show that the evolution of these compact objects is mainly caused by an increase in size that could be explained by the action of dry minor mergers. We detect also an evolution in the Fundamental Plane caused mainly by this population of very compact and bright galaxies. Unfortunately, we cannot distinguish a change in the slope from an increase in the scatter of the Fundamental Plane since our high--redshift sample is biased to the brightest objects.}
   {}

   \titlerunning{Evolution of the Fundamental Plane of 0.2$<$z$<$1.2 Early--type galaxies in the EGS}
   \authorrunning{Fern\'andez Lorenzo et al.}
   \maketitle

%

\section{Introduction}

Elliptical galaxies, mainly composed by old star populations and a small amount of gas and dust, represent an advanced stage of the galaxy formation as well as merging processes, making the study of their properties particularly interesting \citep{2006pnbm.conf..299G}. These galaxies are characterized by a de Vaucouleurs stellar profile, and by an elliptical distribution of their components supported by random motions rather than the ordered rotation of spiral galaxies.

The correlation between the luminosity and the central line--of--sight velocity dispersion known as the Faber--Jackson relation \citep[FJR;][]{1976ApJ...204..668F} relates the stellar and total mass of an elliptical galaxy in a similar way to the Tully--Fisher relation for spirals, and has been used to find relative distances to elliptical galaxies \citep[see for example,][]{1984ApJ...281..512D}. However, while the properties of spirals are rather well described by the two quantities (luminosity and rotation velocity) involved in the Tully--Fisher relation, the large scatter in the FJR forced the authors to look for other parameters to describe the elliptical galaxies. \citet{1977ApJ...218..333K} found that the surface brightness and the effective radius, as defined by de Vaucouleurs, are closely correlated (the Kormendy relation, KR). Subsequently, \citet{1987ApJ...313...59D} and \citet{1987ApJ...313...42D} found that these two parameters plus the central velocity dispersion are the axes of a coordinated system where the galaxies define a plane known as 'Fundamental Plane'(FP). The smaller scatter of the Fundamental Plane provides an improvement factor of 2 over the FJR as a distance estimator. Since the velocity dispersion is preventing the gravitational collapse, it can be used for estimating the galaxy mass by applying the virial theorem. Thus, the study of fundamental relations and their change with redshift can provide valuable information on the dynamical masses and mass evolution of galaxies. Moreover, the FP relates structural properties of galaxies, such as sizes and luminosities with the dynamics, hence its study with redshift has important implications on the formation and evolution of elliptical galaxies.

Local studies show that the FP is followed by both elliptical (E) and lenticular (S0) galaxies \citep{1996MNRAS.280..167J}, which are jointly referred to as early--type galaxies. At higher redshift, the FP of E and S0 galaxies could be different if the star formation activity of S0 galaxies were more intense in the past. However, some studies have compared the results obtained for each type of galaxies and no differences between elliptical and S0 were found \citep[e.g.][]{1996MNRAS.281..985V,1997ApJ...478L..13K}. Nevertheless, \citet{2005MNRAS.358..233F} found a mild and on average larger evolution of lenticular galaxies over ellipticals, investigating 96 galaxies in clusters at z$\sim$0.2.

Pioneering evolutionary studies of the FP were made at intermediate redshift (z$\sim$0.5) for galaxies in clusters. For instance, \citet{1996MNRAS.281..985V} and \citet{1997ApJ...478L..13K} found that the FP at this redshift was similar to the FP for nearby galaxies, suggesting little changes on the structure of the older galaxies and a formation epoch at significantly higher redshift (z$_{\rm form}>$2). These authors also found a lower mass--to--light (M/L) ratio of the galaxies at z$\sim$0.5, consistent with passively evolving stellar populations, but the low number of galaxies in their samples made difficult to quantify the results. In a subsequent work, \citet{2000ApJ...531..184K} increased the sample of galaxies in a cluster at z=0.33 to 53 galaxies, and derived mass--to--light ratios of the high redshift early--type galaxies lower than those in Coma cluster by $\Delta \log \rm (M/L_V)$=-0.13$\pm$0.03. This change in the M/L$_{\rm V}$ ratio would imply an increase of the V luminosity by $\sim$0.3 mag, if such evolution were due to the passive evolution of stellar populations alone. In this sense, \citet{2001ApJ...553L..39V} and \citet{2005A&A...433..519Z} found a similar result from a larger sample of field galaxies, deriving an average increase of the B luminosities by $\sim$$-$0.4 mag at $<$z$>$=0.4. In addition, \citet{2001ApJ...553L..39V} found that the M/L ratios of field galaxies evolves as $\Delta \ln \rm (M/L_B)$=($-$1.35$\pm$0.35)z, similar to cluster galaxies. In contrast, the studies at higher redshift of \citet{2002ApJ...564L..13T} and \citet{2003ApJ...597..239G} found a greater brightening for field galaxies than the one found for cluster galaxies. Restricting the \citet{2003ApJ...597..239G} sample to galaxies at z=0.4, the B luminosities of early--type galaxies would have been $\sim$0.7 mag brighter than nowadays, which was attributed to a more recent formation epoch, around z=1.5.

While all previous authors found an evolution of the FP only in the offset, \citet{2005ApJ...633..174T} find a change in the slope of the FP which can be interpreted as a mass--dependent evolution. Then, massive galaxies evolve passively on longer time--scales, whereas less massive systems have more extended star formation histories and continue to form stars at much later epochs. The recent studies of \citet{2009MNRAS.393.1467F}, for a sample of 24 field galaxies covering redshifts 0.20$<$z$<$0.75, and \citet{2009AN....330..931F}, for a sample of 50 cluster galaxies at z$\sim$1, also support the evolution in the slope of the FP for both field and cluster galaxies, suggesting that the internal properties of a galaxy are more important to its evolutionary history than its enviroment. 

In the present work, we study the evolution of the FP of a sample of 122 early--type field galaxies at $<$z$>$=0.7 in the Extended Groth Strip (EGS). This larger sample of galaxies can provide more statistically significant results than other works, as well as to accurately determine the FP followed by high redshift galaxies. In addition, we analize a sample of 13 local galaxies following the same procedure used for the high redshift sample, to calibrate the local FP used as comparison.

This paper is organized as follows. In Sect. 2, the data and sample selection criteria are presented. Determination of the effective radius, suface brightness, and velocity dispersion are described in Sect. 3. In Sect. 4, the Fundamental Plane is derived and its evolution is analysed using the FJR and the KR. The last two sections provide the discussion and the summary of the results. Throughout this article, the concordance cosmology with ${\Omega}_{\rm \Lambda0}=0.7$, ${\Omega}_{\rm m0}=0.3$ and $\rm H_0=70 \rm \ km \rm \ s^{-1} \rm \ Mpc^{-1}$ is assumed. Unless otherwise specified, all magnitudes are given in the AB system.

\section{Data \& Sample Selection}

The sample consists of galaxies in the EGS sky region. The baseline for spectroscopy target pre--selection were the galaxies for which DEEP2 spectra (Data Release 3, DR3) in this field were available. We selected the objects, from the redshift catalogue, with the tags ZQUALITY=3 or 4 and CLASS=GALAXY, to remove stars and AGNs. The DEEP2 project \citep{2003SPIE.4834..161D,2007ApJ...660L...1D} is a survey using the DEIMOS multi--object spectrograph \citep{2003SPIE.4841.1657F} in the Keck telescope, to study the distant Universe. The grating used was the 1200 l/mm one, covering a spectral range of 6500--9100 {\AA} with a dispersion of 0.33 {\AA}/px, equivalent to a resolution R=$\rm \lambda$/$\rm \Delta\lambda$=4000.

The photometric data used here are part of AEGIS survey \citep{2007ApJ...660L...1D}. B, R, and I--band photometry were taken with the CFH12K mosaic camera \citep{2001ASPC..232..398C}, installed on the 3.6--meter Canada--France--Hawaii telescope (CFHT). These magnitudes are included in the DEEP2 photometric catalogue  \citep[Data Release 1, DR1;][]{2004ApJ...617..765C} and the magnitude errors from sky noise and redshift are available too. The data in the V--band (F606W) were taken from HST catalogue, and obtained using the ACS camera. Although we adopted the I--band data from CFH12K, we used also the images in the I--band (F814W) obtained with the ACS, for deriving the structural properties of our sample of galaxies. In addition, we used the z--band magnitudes of the Canada--France--Hawaii telescope legacy survey (CFHTLS), since z--band at z=1 roughly matches the rest--frame B--band photometry. B, R, and I--band magnitudes are already corrected for Galactic reddening based on \citet{1998ApJ...500..525S} dust maps. The V and z--bands have been corrected following the prescriptions of the same work. At this point, our pre--selected sample has 3862 galaxies with DEEP2 spectrum and B, V, R, I, and z--band magnitudes available.

The second selection criteria was the morphology. In \citet{2009fernandez} we compared the DEEP2 spectra with a synthetic, early--type galaxy spectrum, with the aim of segregating E/S0 from non--E/S0,through the rms of the difference between spectrum and template. We checked the result through a visual classification, and we found that the mismatch for the non--E/S0 group was $\sim$3$\%$, whereas the one associated with the group of E/S0 galaxies was significantly worse, due to contamination by spirals. Then, although we cannot select our sample of E/S0 this way, a preliminary selection using this procedure allowed us to eliminate a large fraction of the sample (2/3), losing only a 3$\%$ of the E/S0 galaxies. For the remaining 1303 galaxies, which are characterized by having a spectra similar to that of the template, we made a visual classification using the summed V+I images of HST/ACS, and we found 400 E/S0 candidates. 

Finally, we selected the E/S0 galaxies that had, at least, one absorption line in the spectrum for determining the velocity dispersion, such as E--band region (Fe line, $\rm \lambda\lambda$5270), G--band region (Fe and Ca lines, $\rm \lambda\lambda$4300), and the $\rm H+K$ region (double CaII line, $\rm \lambda\lambda$3934, 3969). The double absorption line of sodium or the triple absorption line of magnesium were not used since they could not be deblended, due to the lack of spectral resolution.

 \begin{figure*}[!htl]
   \centering
   \includegraphics[angle=0,width=18.5cm]{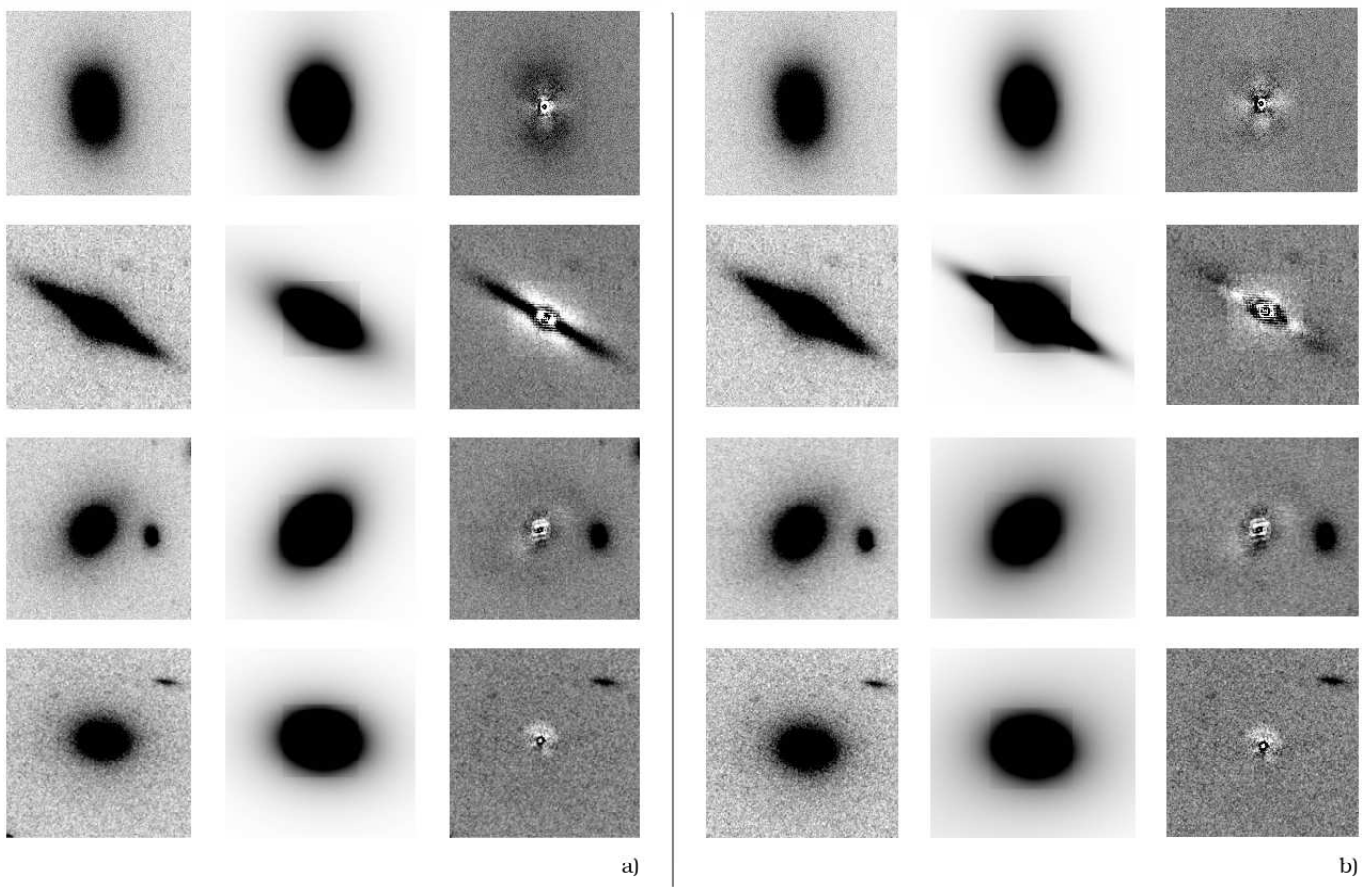}
      \caption{Examples of HST/ACS surface brightness modelling. For each galaxy, we have represented two model fits: a single de Vaucouleurs function (a) and a combination of r$^{1/4}$ + exponential disc component (b). For each galaxy and model we have shown the original data (left panel), the model fit (central panel) and the residual image (right panel).}
   \end{figure*}

\section{Data Analysis}

The final sample of galaxies with photometric information in the B, V, R, I and z--bands, in the redshift range 0.2$<$z$<$1.2 and classified as E/S0, consists of 135 galaxies. The FP is constructed from three of the four fundamental parameters of the E/S0 galaxies: luminosity, velocity dispersion, effective radius and surface brightness at the effective radius. Physical parameters of the galaxies in our sample are shown in table 1 (in electronic format). Here, we describe the way to determine these parameters.

\subsection{Structural parameters}

Historically, the surface brightness and the effective radius of the early--type galaxies have been calculated by fitting the mean radial profiles with a de Vaucoleurs function r$^{1/4}$ \citep{1987ApJ...313...59D,1987ApJ...313...42D}. In more recent evolutionary studies of the FP, other surface brightness profiles such as S\'ersic function, exponential law, or combinations of two types of models \citep[for example,][find that bulge+disc models provide better fits than a pure r$^{1/4}$ law]{2001ApJ...553L..39V} have been used. Although the surface brightness and the effective radius are strongly dependent on the fitting profile, different fits have little effect on the FP, since the galaxies shift along the plane instead of perpendicular to it \citep{2000ApJ...531..184K}.

In the present work, the structural modeling of the galaxies were made on the I--band HST/ACS images, using the {\tt GALFIT} package of \citet{2002AJ....124..266P}. The effects of colour gradients are minimised by using the measurement in the I--band, since it corresponds to an optical rest--frame band in our whole redshift range. {\tt GALFIT} minimises de $\chi^2$ residual between the data image and the model, adjusting free parameters simultaneously, such as the total magnitude or the effective radius. Since different fitting functions are available in this code, and it is possible to deblend objects into one or more components, we explored the results of three profile models: a pure de Vaucouleurs function, a S\'ersic profile, and a combination of r$^{1/4}$ and exponential laws. All models were convolved with a point--spread function (PSF) generated by Tinytim \citep{1995ASPC...77..349K}, and we used the parameters derived by {\tt SExtractor} \citep{1996A&AS..117..393B} in the combined V+I HST images as inputs in the {\tt GALFIT} code. Fig. 1 presents examples of the results of the surface brightness modelling using a single de Vaucouleurs function and a bulge--to--disc decomposition (B/D). The mean difference in surface brightness between both fits is $\sim$0.3 mag. However, the FPs obtained from the structural parameters calculated by these two models are equivalent, according to the above result of \citet{2000ApJ...531..184K}. We have found also the same FP using the structural parameters obtained by fitting a S\'ersic profile.

 \begin{figure*}[!htl]
   
   \includegraphics[angle=0,width=19.5cm]{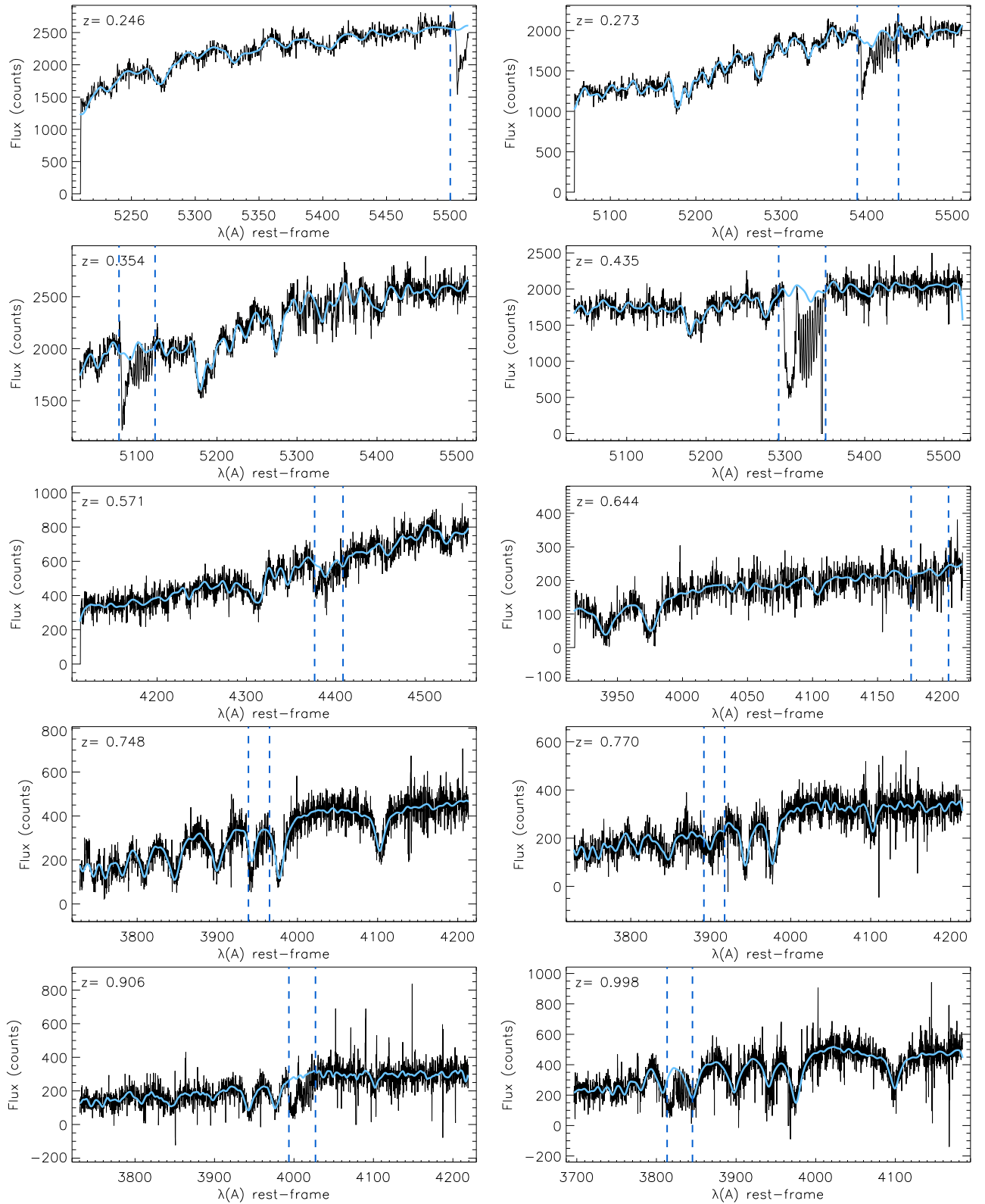}
      \caption{Example spectra (not flux--calibrated) of our sample of galaxies. The thin black line represent the rest--frame spectral data. The thick blue line is the galaxy model spectrum obtained from the convolution between the velocity profile and the stellar templates. The dashed vertical lines demarcate the regions that were excluded from the fitting because of the sky contamination by molecular bands.}
   \end{figure*}

For a proper comparison of galaxies at different redshifts, measured surface brightnesses need to be converted to a common rest--frame band. To calculate these k--corrections, we need to know the spectral energy distribution (SED) of the galaxy. In \citet{2010fernandez} we found that the template obtained by the routine {\tt kcorrect} \citep{2007AJ....133..734B} from a non--negative linear combination of five templates based on the \citet{2003MNRAS.344.1000B} stellar evolution synthesis codes, provides the best result for retrieving the SED of the galaxy. In addition, we found that the most reliable k--correction is obtained from information in an observed band that roughly matches the rest--frame band. Here, we have the magnitudes in the B, V, R, I, and z--bands. Since the z--band roughly matches the B--band at z=1, these five bands provide enough photometric information to calculate the rest--frame B--band magnitudes in our redshift range. Since different instruments have been used to obtain the magnitudes in different wavelengths, the same aperture might not ensure a consistent fraction of light in each band, due to the PSF, the seeing or the pixel scale, that act spreading more or less the object. Then, we compared the k--corrections derived from the B, V, R, I, and z magnitudes with those obtained using the aperture magnitudes (in a 1.5" radius) provided by CFHTLS in the g, r, i, and z--bands, and we found good agreement between both k--corrections. 

The {\tt GALFIT} code provides the effective semimajor axis (a$_{\rm 50}$) rather than the circular effective radius (r$_{\rm e}$). Since local works use structural parameters defined in a circular annuli, we used r$_{\rm e}$=a$_{\rm 50}$$\sqrt{b/a}$, where $b/a$ is the axis ratio, for a proper comparison with other samples. The surface brightness SBe=$-$2.5log$_{10}\Sigma_e$ was determined using \citep{2002AJ....124..266P}

 \begin{equation}
 F_{tot} = 2\pi r_{\rm e}^2\Sigma_e e^k n k^{-2n} \Gamma(2n)b/a \ ,
 \end{equation}
where $F_{tot}$ is the total flux in the rest--frame band (i.e. the flux already k--corrected), $\Sigma_e$ is the surface brightness within the effective radius $r_{\rm e}$, $k$=1.9992$n-$0.3271, $\Gamma$ is the gamma function, $b/a$ is the axis ratio, and $n$ is the S\'ersic index ($n$=4 for the classical de Vaucouleurs profile). The surface brightness was also corrected for cosmological dimming multiplying $F_{tot}$ by (1+z)$^4$.

\subsection{Velocity Dispersion}

The observable, luminous component of early--type galaxies is mostly made of stars. Therefore, their spectra can be described as a sum of individual stellar spectra redshifted according to their line--of--sight velocities. Assuming that the spectrum of all stars is given by a single template, then the galaxy spectra can be described as a convolution between the model template that represents the stellar spectrum and a broadening function that describes the internal kinematics. Early methods for deriving the line--of--sight velocity dispersions (LOSVD) used both the Fourier fitting technique \citep{1989ApJ...344..613F}, where the broadening function is recovered from a deconvolution process, or the correlation method \citep{1979AJ.....84.1511T}, where the galaxy spectrum is cross--correlated against an essentially noiseless stellar template. These techniques are very sensitive to template mismatch and assume a Gaussian form for the broadening function, a simplifying assumption not adequate for the kinematics of elliptical galaxies that frecuently present multi--component stellar structures. These limitations force the researchers to look for better ways to measure the stellar kinematics \citep[see][for an overview of the various methods]{2003MNRAS.339..215D}. For example, \citet{1990A&A...229..441B} derived a new technique ({\tt Fourier correlation quotient, FCQ}), which is a hybrid method that provides a detailed analysis of the shape of the broadening function, and that has been used in recent studies of the FP \citep{2005A&A...433..519Z,2009MNRAS.393.1467F}. Thanks to the increase in the computational speed, the LOSVD has begun to be obtained in the pixel space \citep{1992MNRAS.254..389R,2000ApJ...531..159K}. A treatment of the problem in pixel space has the advantage that gas emission lines or bad pixels can easily be eliminated, the effects of noise are more easily incorporated, and allows an easier error estimate. 

In this work, the velocity dispersion, $\sigma$, has been calculated using the Penalized Pixel--Fitting method ({\tt pPXF}) developed by \citet{2004PASP..116..138C}, that works in the pixel space. This software extracts the stellar kinematics from absorption--line spectra of galaxies, using a maximum penalized likelihood approach. The code creates
a model galaxy spectrum by convolving a template spectrum by a parametrized LOSVD. The template spectrum is a combination of stellar templates with different metallicities and spectral types, and the function describing the LOSVD is based on the Gauss--Hermite series. The parameters of the LOSVD are fitted simultaneously by minimizing the $\chi^2$, which measures the agreement between the model and the observed galaxy spectrum. The maximum penalized likehood formalism, allows obtaining a solution that reproduces the details of the actual profile when the signal--to--noise (S/N) is high, while an adjustable penalty term is added to the $\chi^2$ to bias the solution towards a Gaussian shape when the signal--to--noise (S/N) is low \citep[see][for a complete explanation]{2004PASP..116..138C}. The ability of the {\tt pPXF} method to fit a large set of stellar templates together with the kinematics, allows eliminating the template mismatch problem.

The one--dimensional DEEP2 spectra used in this work, were extracted from the two--dimensional ones by using the routine {\tt extract1d.pro} that is part of the DEIMOS {\tt spec2d} code. We use Horne's optimal extraction algorithm \citep[see][]{1986PASP...98..609H} along the locus of constant lambda. In the spatial direction, we used an aperture equal to the half--light radius determined in the previous section unless that value is lower than 3 ($\sim$PSF) or larger than 10 pixels (the pixel scale of spectra is 0.117 arcsec/pixel). For these extreme cases we adopted 3 and 10 pixels respectively (6 galaxies have an effective radius slightly larger, but the difference in the velocity dispersion is within the errors). Since $\sim$43$\%$ of the sample have a minimum of 3 pixels, we made aperture corrections for these objects (see below). The resulting 1D spectra were corrected to rest--frame wavelength before calling {\tt pPXF}. S/N of spectra have values between 5 and 40, with a median value of 10. In addition, for minimizing the fitting time, we limited the wavelength region used in the analysis from l$_C-$250${\AA}$ to l$_C+$250${\AA}$, where l$_C$ is the absortion line selected in each spectrum for determining the velocity dispersion. In some cases, when the absortion line was at the beginning or the end of the spectral range covered by DEEP2, a shorter wavelength region had to be used during the fit. 

 \begin{figure*}[!htl]
\centering
      \includegraphics[angle=0,width=8.5cm]{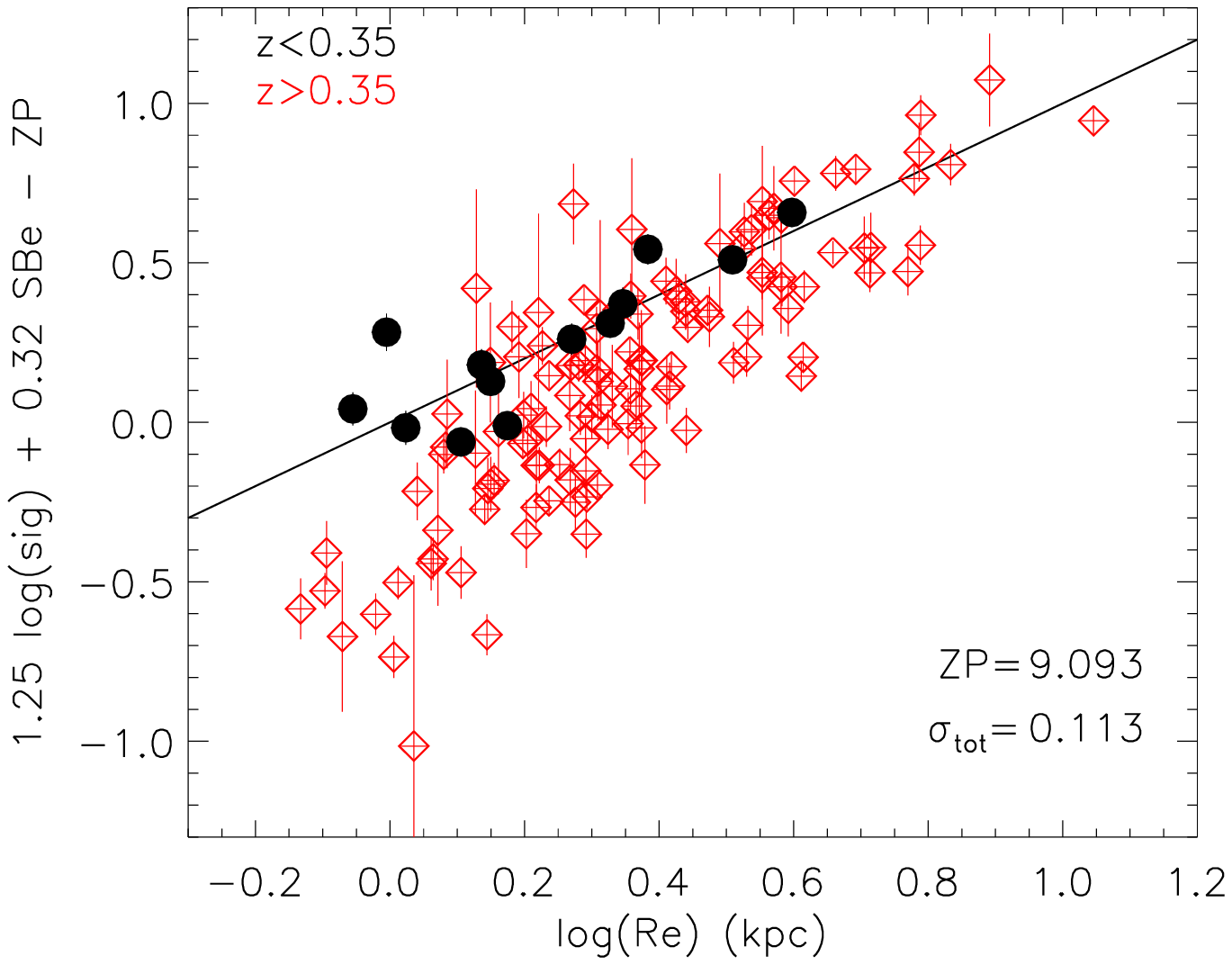}
      \includegraphics[angle=0,width=8.5cm]{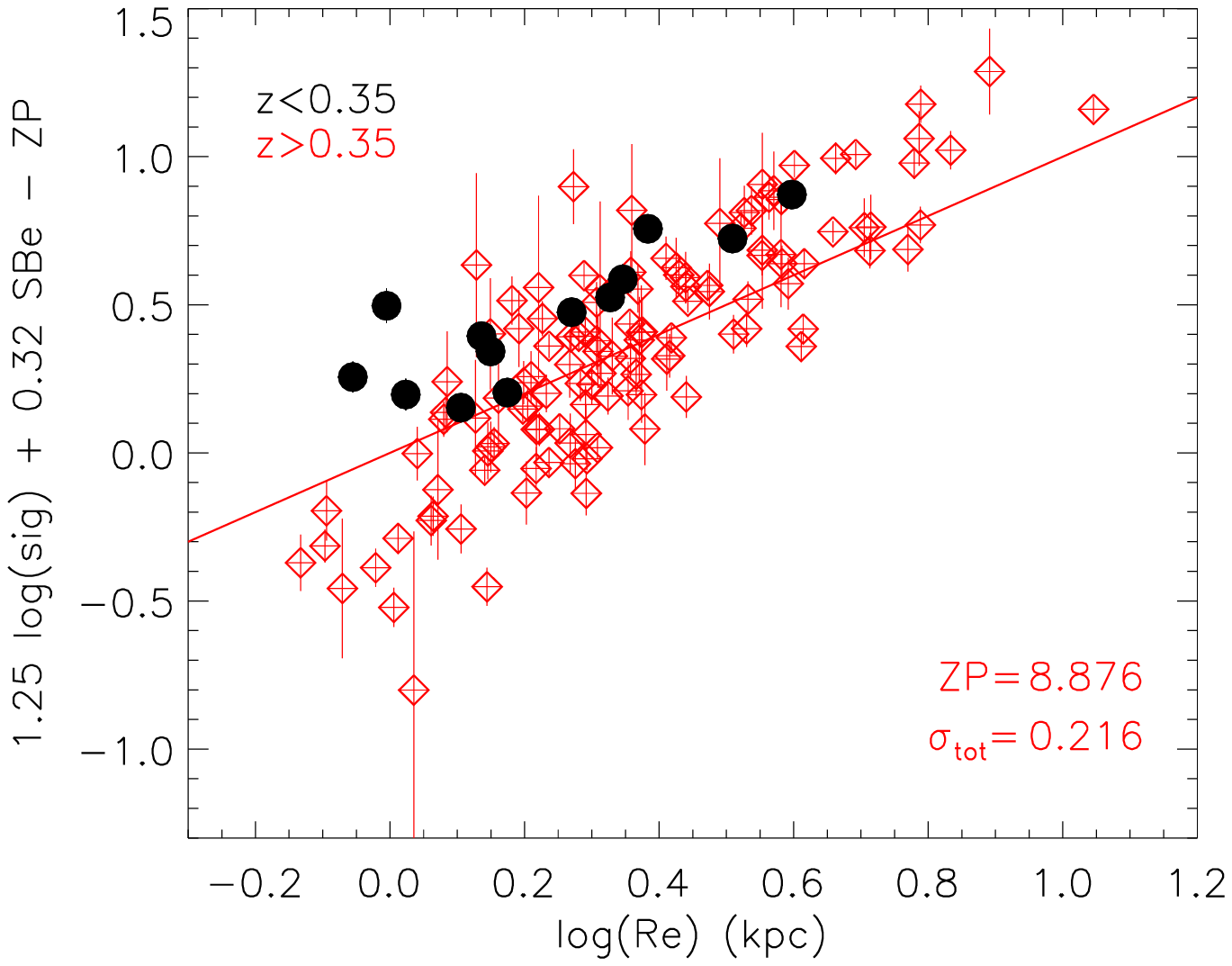}
      \caption{Edge--on projection of the Fundamental Plane in the B--band fitting the galaxies at z$<$0.35 (left) and fitting the data at z$>$0.35 (right). The black points represent the galaxies with z$<$0.35, and the open red diamonds are the objects with z$>$0.35. The galaxies in each redshift range are fitted using the parameters $a$ and $b$ obtained for local galaxies from the literature. The zero--point (ZP) and the rms scatter ($\sigma_{tot}$) are shown in both cases.}
   \end{figure*}

 \begin{figure*}[!htl]
\centering
      \includegraphics[angle=0,width=8.5cm]{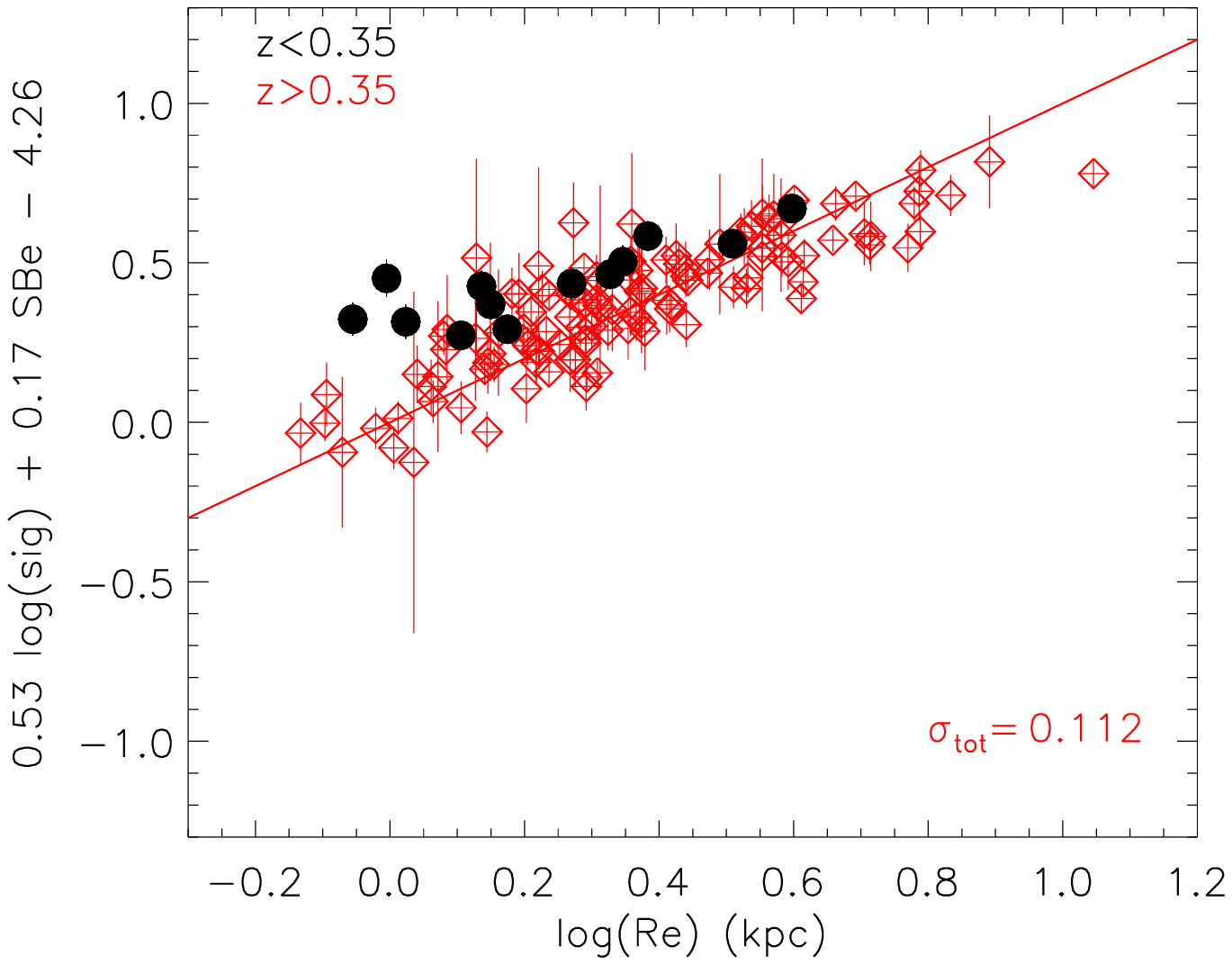}
      \includegraphics[angle=0,width=8.5cm]{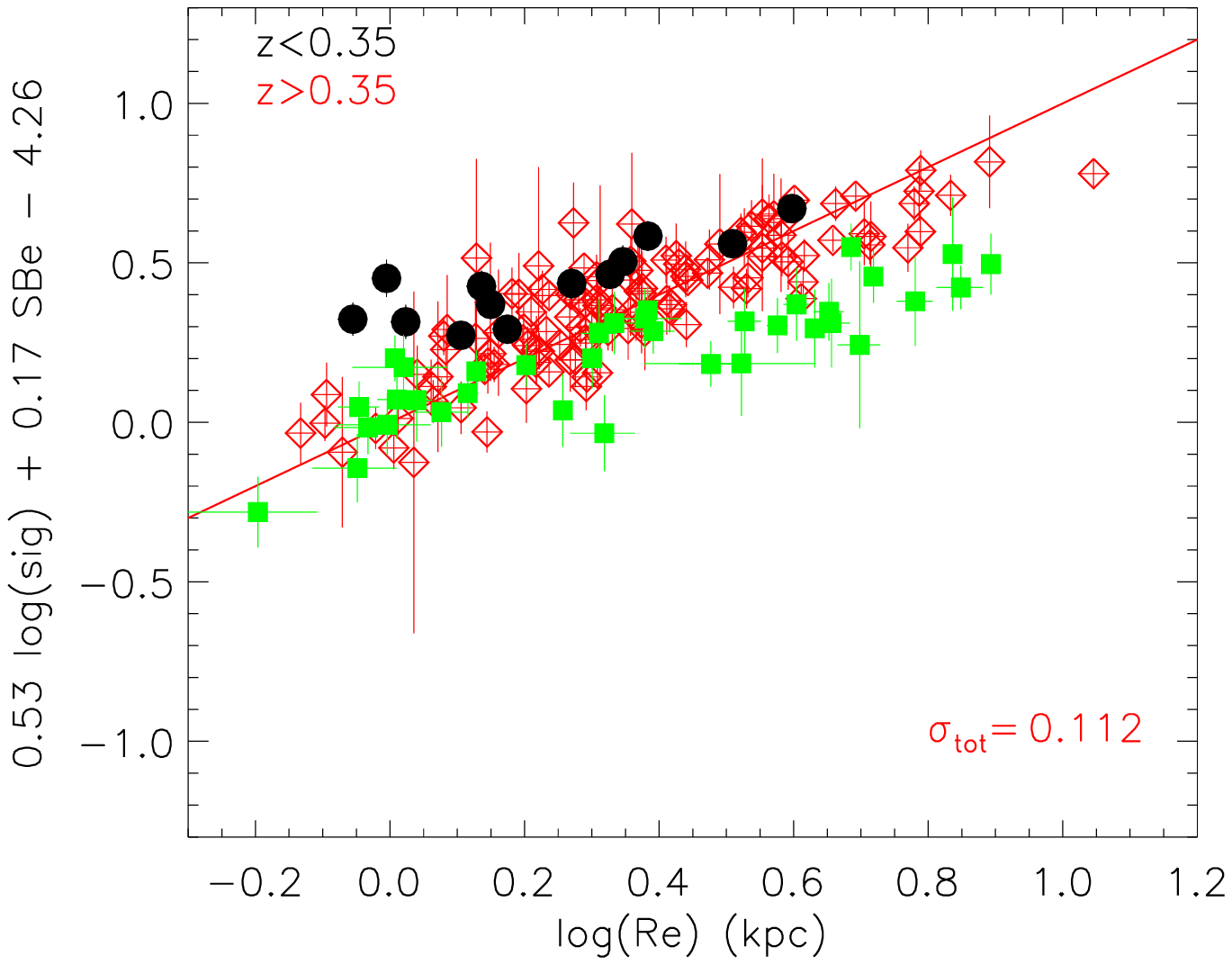}
      \caption{Edge--on projection of the Fundamental Plane in the B--band. The black points represent the galaxies with z$<$0.35, and the open red diamonds are the objects with z$>$0.35. The parameters $a$, $b$, and the zero--point were obtained by fitting the galaxies at z$>$0.35, through a non--linear weighted least squares fit. The green squares added in the right panel are the early--type galaxies of \citet{2003ApJ...597..239G} with redshifts in the range 0.3$<$z$<$1.0.}
   \end{figure*}

To determine the velocity dispersion for our wide redshift range (0.2$<$z$<$1.2), we need to use different absorption lines, such as E--band region (Fe line, $\rm \lambda\lambda$5270), G--band region (Fe and Ca lines, $\rm \lambda\lambda$4300), and the $\rm H+K$ region (double CaII line, $\rm \lambda\lambda$3934, 3969). The $\rm H+K$ region is, a priori, not very suitable for measuring the velocity dispersion because the CaII lines are intrinsically broad. Since the $\rm H+K$ region is the most accessible kinematic diagnostic in galaxies at z$\sim$1, \citet{2000AJ....119.1608K} investigated the velocity widths measured from the $\rm H+K$ region and from the HI line (21cm), for a sample of late--type galaxies. They found a good correlation between both measures, by using a maximum--penalized likelihood approach and a set of stellar templates of 12 Galactic A main--sequence through K giant stars. In addition, \citet{2003ApJ...597..239G} obtained good agreement between the internal kinematic measured separately over the G and the $\rm H+K$ regions, using the same procedure for a sample of early--type galaxies. Nevertheless, choosing an appropriate set of stellar templates is of the utmost importance to obtain a reliable kinematic from the $\rm H+K$ region \citep{2000AJ....119.1608K}. The stellar templates used in this work have been taken from the Indo--U.S. Library of Coud\'e Feed Stellar Spectra \citep{2004ApJS..152..251V}, which cover a spectral range from 3460 to 9464 $\AA$, at a resolution of $\sim$1 $\AA$ FWHM and at an original dispersion of 0.44 $\AA$ pixel$^{-1}$. From this library, we have selected a subsample of 18 stars with different metallicities and spectral types ranging from A to K. For a proper comparison with the DEEP2 spectra, each stellar template is convolved with the quadratic difference between the DEEP2 and the Coud\'e Feed Stellar Spectra instrumental resolution. Finally, we excluded some regions from the fit because of the contamination by molecular bands of the sky. In Fig.2 we have represented some examples of DEEP2 spectra, the best galaxy model fitted by {\tt pPXF}, and the regions excluded from the fit in some cases.

We use Monte Carlo simulations to estimate measurement errors for the LOSVD extraction method. With this aim, we added noise to the original data and repeated the full measurement process for 100 realizations. Finally, we calculated the standard deviation from the distribution of values given by the simulation and used them as velocity errors. We found that $\sim$90$\%$ of the objects in our sample had errors lower than 10$\%$. 

Since we have used spectra extracted from physical sizes that not correspond to the effective radius in some cases, the velocity dispersion of these objects must be aperture corrected. We have followed the method of \citet{2003A&A...407..423M}, who obtained the aperture corrections by computing velocity dispersion gradients. The median value of the $\sigma$--gradient measured by these authors is slightly steeper than the one proposed by \citet{1995MNRAS.276.1341J}, but consistent within the error. Moreover, the difference in the aperture correction obtained from both methods is lower than 1 km/s for all objects in our sample.
\clearpage
\section{Results}

\subsection{The Fundamental Plane}

The FP relates the effective radius r$_{\rm e}$, the central velocity dispersion $\sigma$, and the surface brightness within the effective radius $\rm SB_e$ in the following form:

\begin{equation}
\log r_{\rm e} = a \log \sigma+b \ SB_e+c
\end{equation}

To derive the evolution of the FP we have divided our sample in two redshift ranges: galaxies at z$<$0.35, with a mean value of 0.27, which can be considered as local galaxies since they follow a relation that is similar to that found in nearby clusters \citep{2000ApJ...531..184K}; and a high--redshift sample, which is made of galaxies at z$>$0.35, with a mean value of 0.68. The local sample is composed only of 13 galaxies. This number of galaxies is insufficient for determining the local FP parameters but it can be used to calibrate the local FP used as comparison to the high--redshift galaxies, since both samples were analysed following the same procedure. We adopt then, for the local coefficients, the same values used by \citet{2003ApJ...597..239G} and \citet{2009MNRAS.393.1467F} in the rest--frame B--band: $a=$1.25 and $b=$0.32, and we calculated the zeropoint $c$ for the local and high--redshift samples, by using a non--linear weighted least squares fit. These $a$ and $b$ values were obtained by these authors from the local samples of \citet{1989ApJS...69..763F} and \citet{1993A&A...279...75S} respectively, using the same cosmology that we adopted in the present work. In Fig. 3, we present the edge--on projection of the FP in the local (left) and high--redshift samples (right), in addition to the zeropoint and total rms scatter obtained in each case. For our local sample, we obtained a zeropoint of $c=-$9.093, in good agreement with the previous result of \citet{2003ApJ...597..239G} ($-$9.062); and the resulting rms scatter $\sigma_{tot}=$0.113 is not significantly higher than the local value \citep[$\sim$0.1 dex;][]{2005MNRAS.358..233F}. However, we found a different FP intercept for the high--redshift galaxies, according to previous studies of FP evolution \citep[for example][]{2003ApJ...597..239G, 2009AN....330..931F}. This difference in the zero--point is usually interpreted as a difference in surface brightness caused by luminosity evolution, under the assumption that all early--type galaxies evolve in the same way, i.e. the coefficients $a$ and $b$ are independent of the redshift. From the change in the zeropoint of the FP we found a brightening of 0.68 mag in the B--band for early--type galaxies at $<$z$>$=0.7. Using the expression of \citet{2001ApJ...553...90V} for the evolution of a single--age stellar population (L$\propto$1/(t$-$t$_{form}$)$^{k_B}$) and the same value of $k_B$=0.91 used by these authors, this evolution of $\Delta$M$_B$=$-$0.68 mag implies a formation redshift higher than z=5. This evolution could be underestimated in the frame mark of progenitor bias effect. At z=0, we are observing early-type galaxies that could come from merged spirals at high--z, unobserved in high--redshift samples. Therefore, the high redshift sample is a biased subset of the low redshift one. \citet{2001ApJ...553...90V} studied the observed and true evolution of $\log (M/L_B)$ and found a difference of $\Delta \log \rm (M/L_B)$$\sim$0.1  at $<$z$>$=0.7. Assuming that this difference is only due to luminosity, the true absolute magnitude is $\sim$0.25 mag brighter than the observed one. Then, the luminosity evolution found in this work from the FP is underestimated. Nevertheless, the rms scatter obtained in the high--redshift sample under this assumption is double than the local value. Then, either there is an evolution in the FP rms scatter, or the high--redshift galaxies follow a FP that is tilted with respect to the local one.

To investigate the last possibility, we recalculated the FP coefficients without any previous assumption. The result is presented in Fig. 4. The resulting plane is tilted with respect to the local one, and it has a rms scatter of $\sigma_{tot}=$0.112, in good agreement with the local value. In addition, this high--redshift FP presents a rms scatter which is half of that obtained using the local $a$ and $b$ coefficients. Since our high--redshift sample consists of a large number of objects, this result is statistically significant. To check these high--redshift coefficients of the FP, we have represented the data of \citet{2003ApJ...597..239G} (the redshift range is similar to that used in the present work), together with ours, in Fig. 4 (right). We have recalibrated the data of \citet{2003ApJ...597..239G} to the AB system, for comparison. The objects with log(R$_{\rm e}$)$<$0.5 are consistent with the high--redshift FP, but the larger galaxies (log(R$_{\rm e}$)$>$0.5) have a lower zero--point than the similar objects in our sample. However, the rms scatter of the \citet{2003ApJ...597..239G} data for the whole sample is also substantially reduced when they are fitted using $a$=0.53 and $b$=0.17. 

The change of the FP with redshift seems to indicate a different evolution of early--type galaxies according to their intrisic properties, such as total mass, size or luminosity. In the next sections we analise the FJR and the KR to clarify which of these properties is responsible for the change found in the present work. We also study the bias caused by the limited luminosity of the high--redshift sample, that can be affecting the results.

\subsection{The Faber--Jackson relation}

 \begin{figure}
\centering
      \includegraphics[angle=0,width=9.cm]{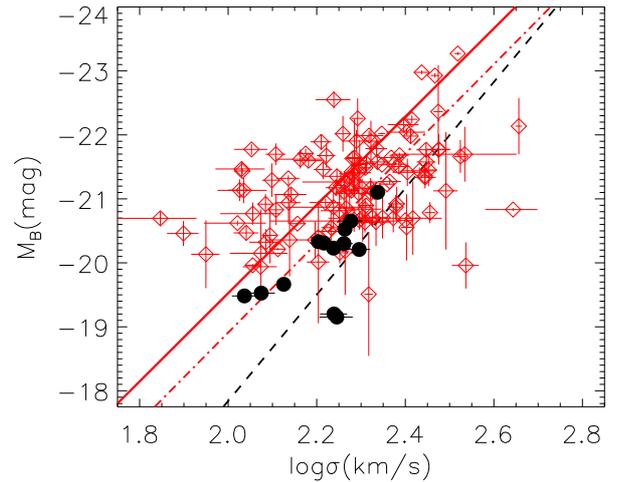}
      \caption{Faber--Jackson relation for the early--type galaxies in our sample. The black points represent the galaxies with z$<$0.35, and the open red diamonds are the objects with z$>$0.35. The dashed black line is the local relation used by \citet{2009MNRAS.393.1467F}, and the dot--dashed red line is the fit to their distant sample (0.2$<$z$<$0.75). The solid red line is the weighted fit to our high--redshift sample.}
   \end{figure}

\begin{table*}
\caption{Parameters of the FJ and Kormendy relations obtained by fitting DEEP2 data in the redshift range z$>$0.35. In addition, we show the local relations derived by \citet{2009MNRAS.393.1467F} from the Coma cluster galaxies of \citet{1993A&A...279...75S}.}             
\centering                          
\begin{tabular}{c c c c c c c}
\hline\hline
\multicolumn{1}{c}{}&\multicolumn{3}{c}{\bf Faber--Jackson relation}&\multicolumn{3}{c}{\bf Kormendy relation}\\
Redshift range & intercept & slope & ${\sigma}_{total}$ & intercept & slope & ${\sigma}_{total}$ \\\hline
Local relation of \citet{2009MNRAS.393.1467F} & -1.22 & -8.31 &  & 19.95 & 2.18 &   \\
z$>$0.35 DEEP2 galaxies & $-5.730\pm0.033$ & $-6.897\pm0.014$ & 0.819 & $18.553\pm0.033$ & $4.142\pm0.073$ & 0.586 \\
\hline
\end{tabular}
\end{table*}

The FJR links the luminosity to the velocity dispersion of galaxies. Assuming that the velocity dispersion is related to the total mass of the galaxy, the FJR allows us to study the evolution in luminosity with the mass for a given galaxy. In Fig.5, we represented the FJR in the B--band for all the galaxies in our sample. As comparison, we show the local relation obtained by \citet{2009MNRAS.393.1467F} from the Coma galaxies of \citet{1993A&A...279...75S} (determined using the same cosmological assumptions that have been adopted in the present work), as well as the fit to the distant sample of \citet{2009MNRAS.393.1467F} (0.2$<$z$<$0.75). We have recalibrated our data to the Vega system for comparison with the work of \citet{2009MNRAS.393.1467F}. The galaxies in our local sample are in good agreement with the local relation, although our objects have in average lower velocity dispersions. However, for the high--redshift sample, we found that the galaxies are brighter than their local counterparts for a given velocity dispersion. In table 2, we present the parameters of the FJ relation obtained by fitting DEEP2 data in the redshift range z$>$0.35 and its total scatter, together with the local relation derived by \citet{2009MNRAS.393.1467F}. We found evidence of a different evolution of the galaxies as function of their masses. Then, although all the galaxies were brighter in the past, we obtained a stronger evolution for the lower mass galaxies. Nevertheless, the large dispersion of this relation makes difficult to obtain a reliable conclusion on the slope change of the FJR. This result is similar to that found by \citet{2009MNRAS.393.1467F}, but the average evolution in the B--band magnitude is more noticeable in our high--redshift sample ($\Delta \rm M_B \sim -$1mag), probably because our average redshift is larger than theirs.

\subsection{The Kormendy relation}

The Kormendy relation (KR) between the surface brightness and the effective radius and its change with the redshift, allows us to study the evolution in luminosity as function of the galaxy size.

The KR in the rest--frame B--band for our local and high--redshift samples of galaxies is represented in Fig.6. As in the previous sections, we used the local relation derived by \citet{2009MNRAS.393.1467F} from the local sample of \citet{1993A&A...279...75S}, as comparison, showing good agreement with our local sample (as in the previous section, we have recalibrated our data to the Vega system). Although the weighted fit to our local sample provides a steeper relation than those of \citet{2009MNRAS.393.1467F}, this discrepancy may be attributed to a single galaxy with a very weak surface brightness (less than 22 mag). For the high--redshift sample, we found evolution in the surface brightness for a given effective radius, in the sense that galaxies were brighter in the past, with a change more noticeable for the smaller galaxies. In table 2, we also show the parameters of the KR obtained by fitting DEEP2 data in the redshift range z$>$0.35, together with the local relation derived by \citet{2009MNRAS.393.1467F}. Similar results have been recently found for higher redshift galaxies by \citet{2009ApJ...705..255T} and \citet{2009ApJ...695..101D} in the i and r--bands respectively. However, the variation in the slope of the KR is attributed by these authors to selection effects, because the sample is biased to the brightest galaxies. Since we have the same selection effects, and our high--redshift galaxies have absolute magnitudes brigher than the local ones, we need a local sample covering a wide range of luminosity for comparison with our high--redshift galaxies.

 \begin{figure}
\centering
      \includegraphics[angle=0,width=9.cm]{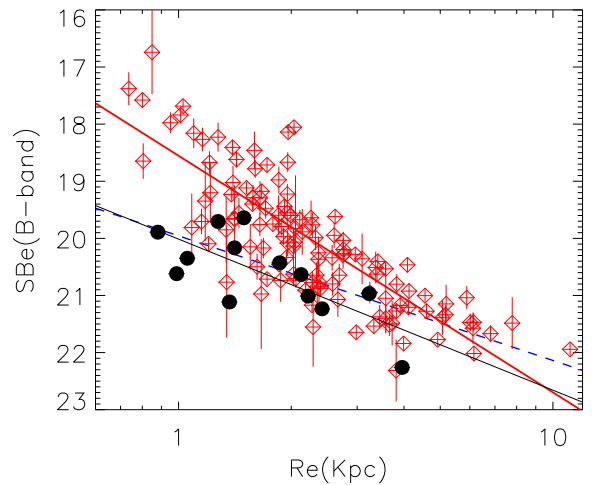}
      \caption{Kormendy relation for the early--type galaxies in our sample. The black points represent the galaxies with z$<$0.35, and the open red diamonds are the objects with z$>$0.35. The dashed blue line is the local relation used by \citet{2009MNRAS.393.1467F}, the thin black line is the weighted fit to our local sample, and the thick red line is the weighted fit to our high--redshift sample.}
   \end{figure}

The largest sample of local galaxies is likely the Sloan Digital Sky Survey database (SDSS). \citet{2003AJ....125.1866B} derived the FP relations in the g, r, i, and z--bands from a magnitude--limited sample of nearly 9000 early--type galaxies in the SDSS, covering a redshift range of 0.01$<$z$<$0.3. The structural parameters of the SDSS sample were obtained by fitting a de Vaucouleurs model to the observed surface brightness profile that accounts for the effects of seeing \citep{2003AJ....125.1866B}. For comparison with our high--redshift sample, we used the surface brightnesses and the effective radii obtained by these authors only in the g--band. Then, we need to calculate the rest--frame g--band magnitudes for our sample of galaxies. To this purpose, we used the k--corrections derived from the aperture magnitudes (in a 1.5" radius) of CFHTLS in the g, r, i, and z--bands (see Sect. 3.1), Finally, we determined the surface brightness within the effective radius in the g--band using equation 1. In Fig.7 we present the comparison of the absolute magnitudes distribution in the g--band for the SDSS sample of \citet{2003AJ....125.1866B} and for our sample of galaxies. The absolute magnitudes of the SDSS galaxies were calculated from the surface brightness and the effective radius, by using equation 1. Both distributions are very similar, showing that the sample of galaxies in EGS that have DEEP2 spectra, is appropriately selected as representative of the whole field, and then both samples are comparable. In the upper panel of Fig. 8, we represented the KR in the g--band for our sample of galaxies together with the Sloan data. Again, we found good agreement for the galaxies in the local sample. Note that the \citet{2003AJ....125.1866B} sample is magnitude limited and therefore the fainter objects in our local sample, that correspond to the galaxies with the highest surface brightness and the lowest effective radius, are rather scarce. On the other hand, the evolution found previously for the high--redshift sample is not so clear here. For a proper comparison, we limited our high--redshift sample and the SDSS one to the same luminosity range, $-$21.5$>$M$_g$$>$$-$22.5, which roughly corresponds to $-$21.0$>$M$_B$$>$$-$22.0. We represented the results in the bottom panel of Fig.8. The largest objects in our high--redshift sample are in good agreement with the local ones, and therefore no evolution in size or luminosity is found for these objects. However, we have a population of objects with low effective radius and high surface brightness that is not present in the local sample. We checked the profile fitting of these $``$small size$"$ galaxies when we found this population, almost inexistent in the local universe. The fit seems good in all cases, with a very low $\chi^2$. The R$_{\rm e}$ is between 4 and 10 pixels (semimajor axis between 5 and 15 pixels) and the values of b/a are distributed along the whole range (0.3--1). The objects are at least 2 times larger than the mean PSF of the ACS band used, and more or less half of them are ellipticals and the other half are lenticulars. We calculated the number of objects with R$_{\rm e}$$<$2 Kpc and luminosities $-$21.5$>$M$_g$$>$$-$22.5 that exist in the comoving volume corresponding to each sample. In the SDSS sample of \citet{2003AJ....125.1866B} (V$_{\rm p}$=0.26 Gpc$^3$), we found $\sim$96 Obj/Gpc$^3$, while the result for our high--redshift sample (V$_{\rm p}$=0.001 Gpc$^3$) is $\sim$24000 Obj/Gpc$^3$, i.e. only the 0.4$\%$ of these objects with R$_{\rm e}$$<$2 Kpc and total luminosity $-$21.5$>$M$_g$$>$$-$22.5 exist in the local universe. Then, an evolution in luminosity or size since z=1 is necessary to explain the decrease in number of these bright and compact objects.

 \begin{figure}
\centering
      \includegraphics[angle=0,width=9.cm]{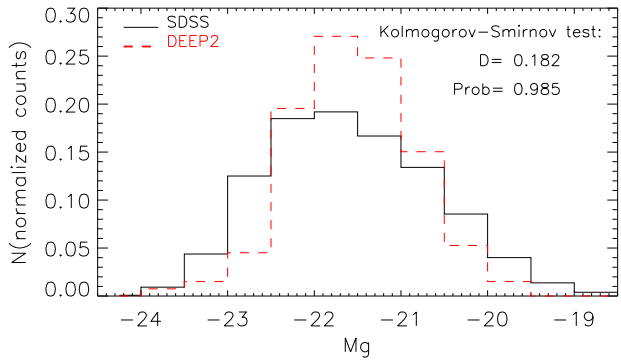}
      \caption{Absolute magnitudes distribution in the g--band of all DEEP2 galaxies in our sample (dashed red line) and of galaxies in the SDSS sample of \citet{2003AJ....125.1866B} (solid black line).}
   \end{figure}

 \begin{figure}[!htl]
\centering
      \includegraphics[angle=0,width=9.cm]{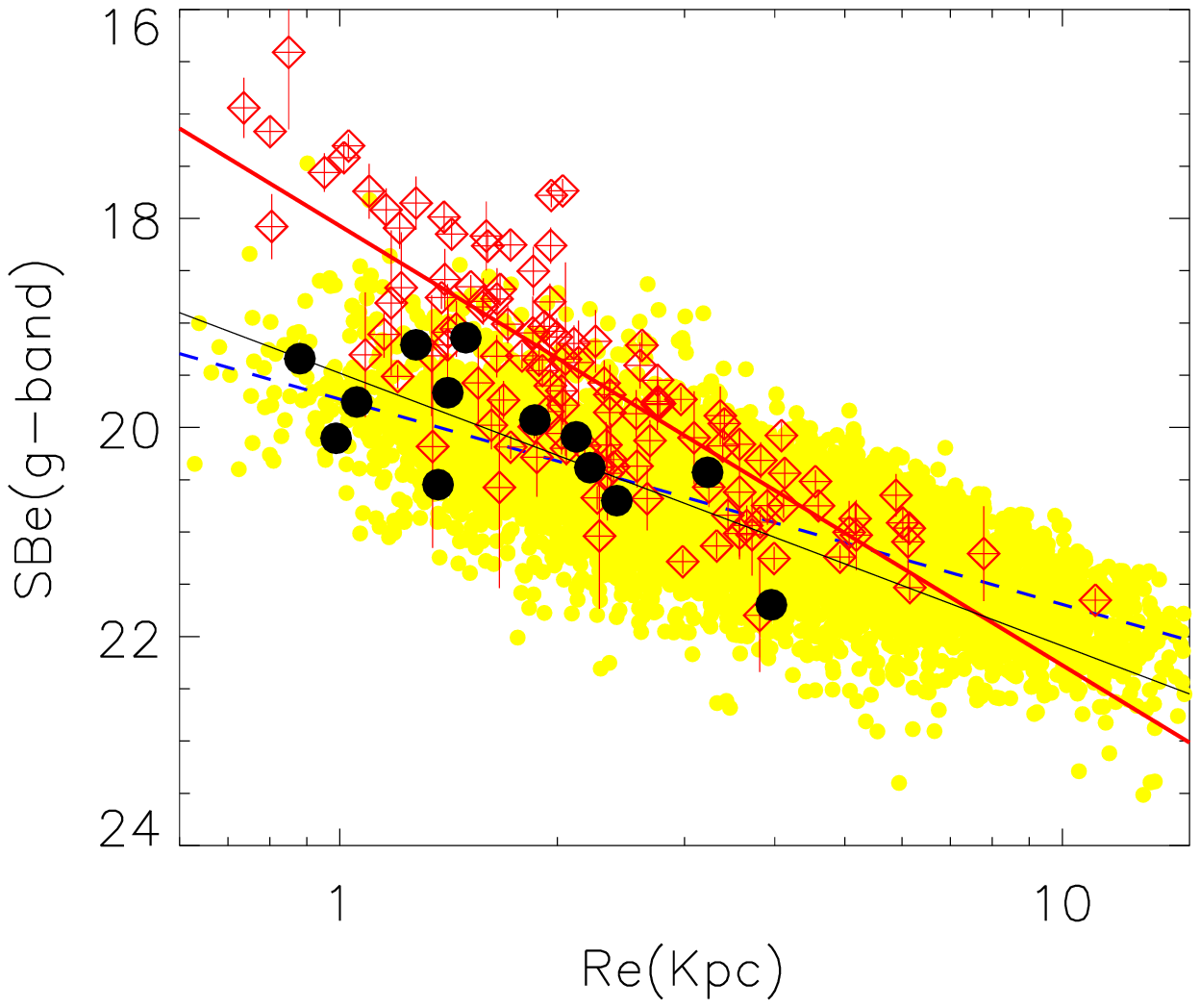}
      \includegraphics[angle=0,width=9.cm]{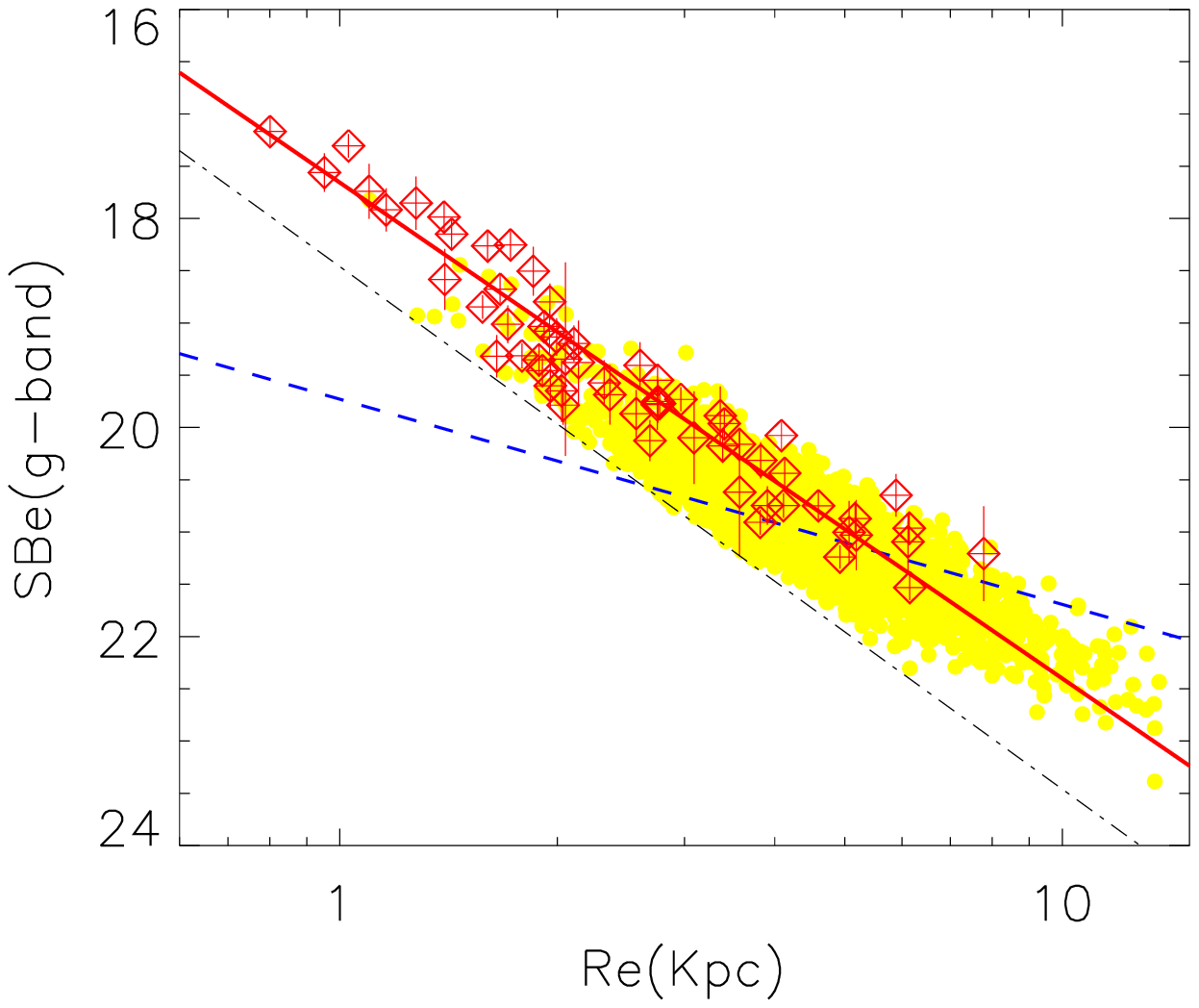}
      \caption{Upper panel: Kormendy relation in the g--band for the early--type galaxies in our sample. The black points represent the galaxies with z$<$0.35, and the open red diamonds are the objects with z$>$0.35. The small yellow points represent the early--type galaxies of SDSS \citep{2003AJ....125.1866B}, and the dashed blue line is the lineal fit to this whole sample. The thin black line represents the weighted fit to our local sample, and the thick red line is the weighted fit to our high--redshift sample. In the bottom panel we have represented the same as for the upper panel, but limiting our high--redshift sample and the SDSS one to the objects with absolute magnitudes in the range $-$21.5$>$M$_g$$>$$-$22.5. The dot--dashed line correspond to M$_g$=$-$21.5.}
   \end{figure}

 \begin{figure*}[htl!]
\centering
      \includegraphics[angle=0,width=18.8cm]{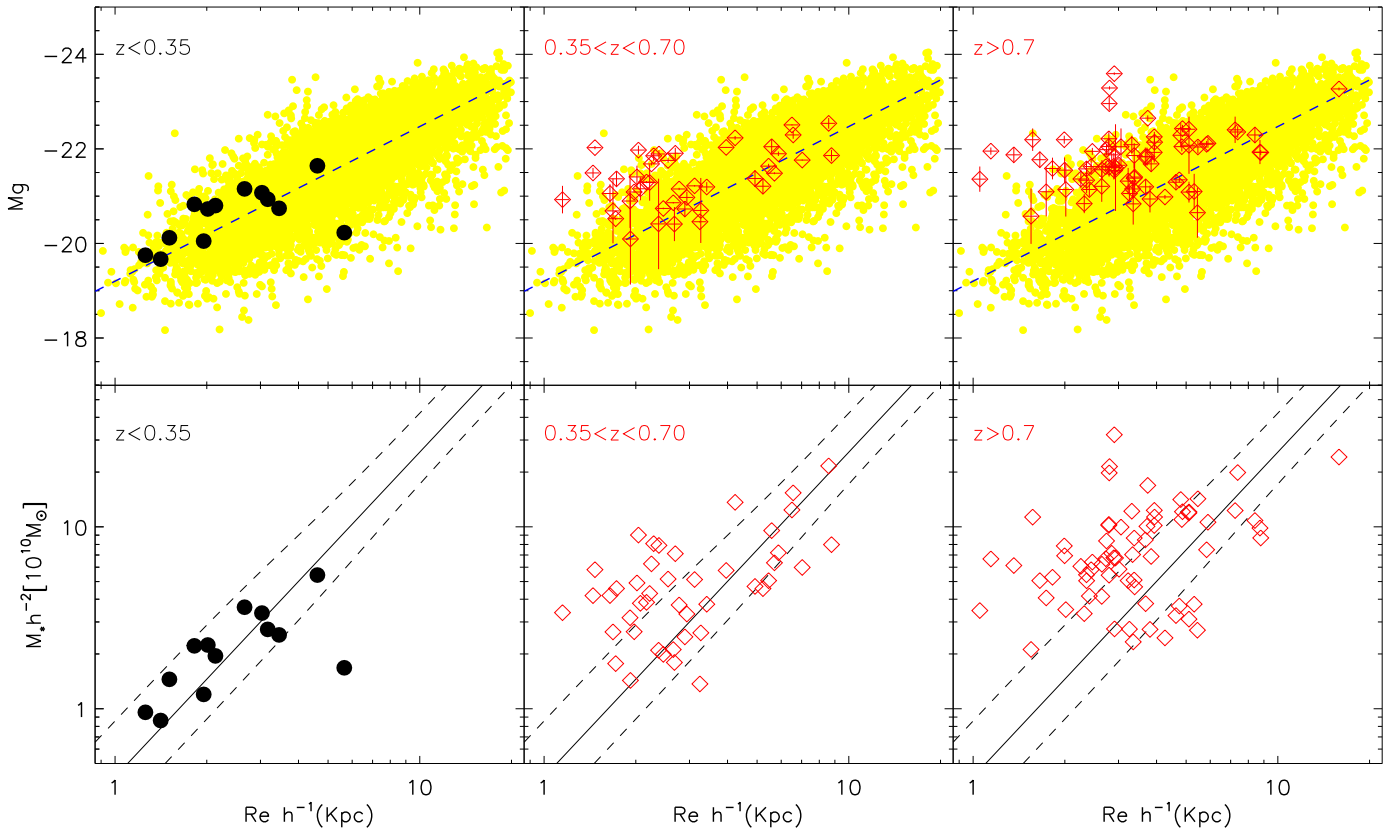}
      \caption{Luminosity--size relation in the g--band (up) and stellar mass--size relation (bottom) for our sample of early--type galaxies. The black points represent the galaxies with z$<$0.35, and the open red diamonds are the objects with z$>$0.35, divided in two redshift ranges. The small yellow points are the early--type galaxies from SDSS \citep{2003AJ....125.1866B} and the dashed line is the linear fit to these data. The solid and dashed lines in the bottom panels represent the stellar masses of \citet{2003MNRAS.343..978S}, as function of the local size distribution of early--type galaxies and its 1 $\sigma$.}
   \end{figure*}

\section{Discussion}

In order to study the processes driving early--type galaxy evolution at z$<$1, \citet{2005ApJ...632..191M} analysed the luminosity--size and stellar mass--size relations in the redshift range 0$<$z$<$1. They found evolution in the luminosity--size relation that is consistent with passive aging of the galaxy stellar population, with the largest evolution appearing for the smallest galaxies. In addition, they ruled out any substantial evolution in the stellar mass--size relation for the same early--type galaxies, consistently with the passive evolution scenario. However, the evolution in the stellar mass--size relation has recently been stablished at higher redshifts \citep{2007MNRAS.382..109T,2008ApJ...687L..61B} and the combination of Palomar massive galaxies sample and lower stellar masses galaxies from the GEMS survey conducted by \citet{2007MNRAS.382..109T} has shown a perceptible evolution in the stellar mass--size relation of early--type galaxies since z$\sim$0.65.

In Fig.9 we present the luminosity--size relation in the g--band (upper panels) and the stellar mass--size relation (bottom panels) for our sample of galaxies. The current surviving stellar mass of our galaxies is calculated by {\tt kcorrect} as the model mass derived from the coefficients fit to each template. The stellar masses used in the present work are then assuming the Chabrier stellar initial mass function. We used the SDSS data of \citet{2003AJ....125.1866B} as local comparison of the luminosity--size relation. For the local stellar mass--size relation, we overplotted the distribution in the stellar mass of SDSS early--type galaxies of \citet{2003MNRAS.343..978S} as function of the S\'ersic half--light radius. We found that the largest objects (R$_e$ larger than $\sim$3 Kpc) follow very similar relations than their local counterparts. However, the smaller objects are on average brighter and have larger stellar masses than the z=0 objects with similar sizes. The same result is obtained using the S\'ersic half--light radius determined by {\tt GALFIT} for our sample of galaxies instead of the de Vaucouleurs one. Although a passive aging of the galaxy population, combined with different formation epochs, can explain the evolution in the luminosity--size relation, the change in the stellar mass is not compatible with this explanation. The evolution of these compact objects is then mainly driven by an increase in size. \citet{2010ApJ...709.1018V} studied the growth of massive galaxies since z=2 and found that it is due to a gradual buildup of their outer envelopes. These authors found that the mass in the central regions is roughly constant with redshift, whereas the growth of the outerparts is due to a combination of star formation and mergers. Also, that star formation is important at the highest redshifts, but the galaxy growth is domitated by mergers at z$<$1, which is expected in $\Lambda$CDM galaxy formation models. Gas--rich mergers increase the star formation activity and cannot explain the old stellar ages of early--type galaxies. However, dry (dissipationless) mergers are more efficient increasing the size than the stellar mass, and therefore would be the dominant mechanisms of size and stellar mass growing. In addition, \citet{naab2009} performed a hydrodynamic simulation and showed that the mass assembly histories dominated by minor ($``$dry$"$) mergers and accretion of stars can explain the increase in size whereas major mergers cannot. 

On the other hand, a decrease in velocity dispersion with redshift is obtained by \citet{2010ApJ...709.1018V} and \citet{cenarro2009}, in contrast to our apparent increase (see Fig.5). However, no evolution is found in the FJR if we compare our sample of galaxies and the local data of \citet{2003AJ....125.1866B} due to the large scatter of the local relation. Since our velocity dispersion is measured only in the central part of the galaxy, and this region is roughly constant with redshift, our result of no change in the velocity dispersion is consistent with this scenario of $``$dry$"$ minor mergers that act increasing the size of the galaxies, due to a buildup of their outer envelopes. In addition, for galaxies with similar masses and luminosities, this change in size would be more noticeable for the most compact and small objects, while the growth in size for the largest objects would remain undetected.

\subsection{Effects of compact objects on the Fundamental Plane}

 \begin{figure*}
\centering
      \includegraphics[angle=0,width=9.cm]{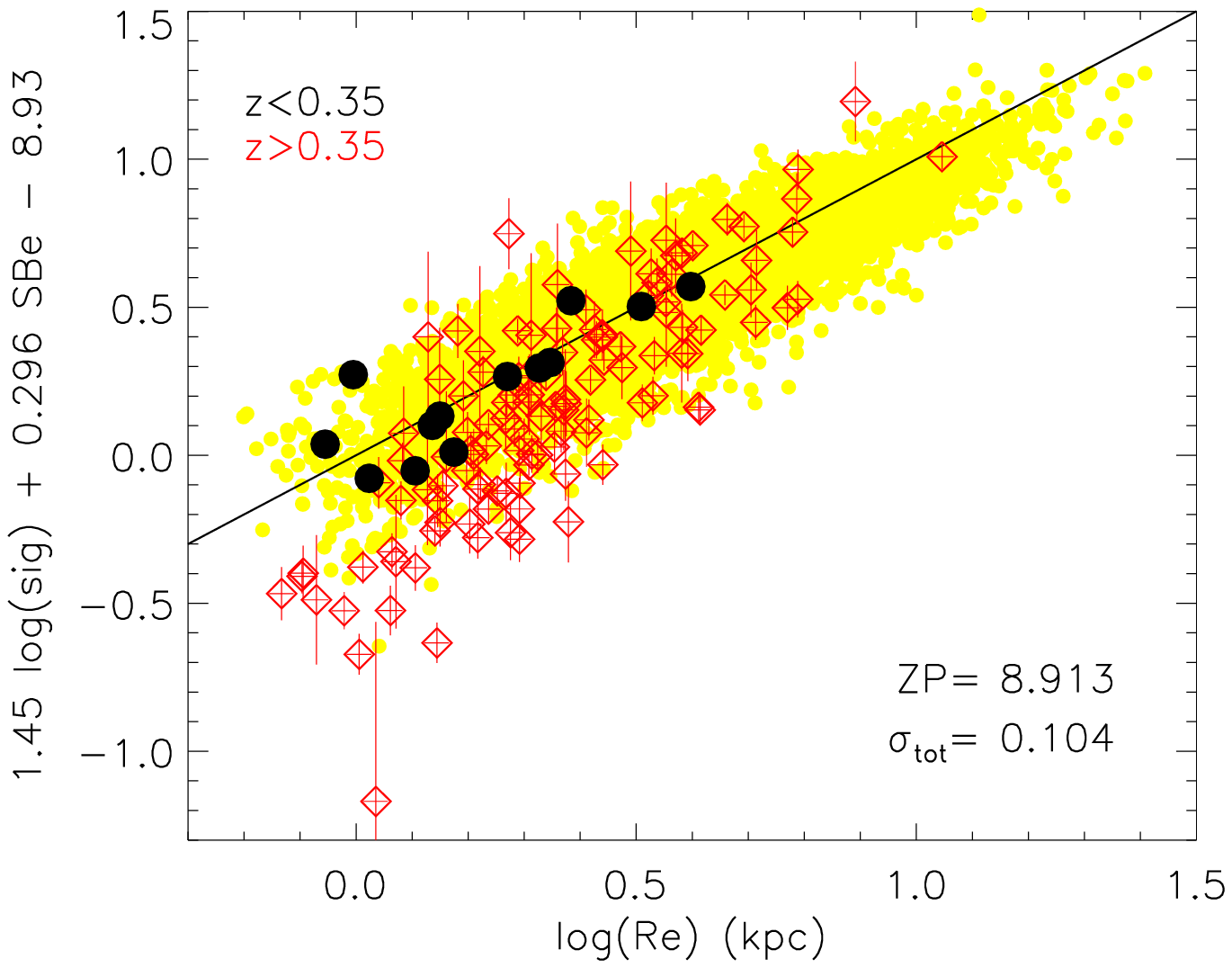}
      \includegraphics[angle=0,width=9.cm]{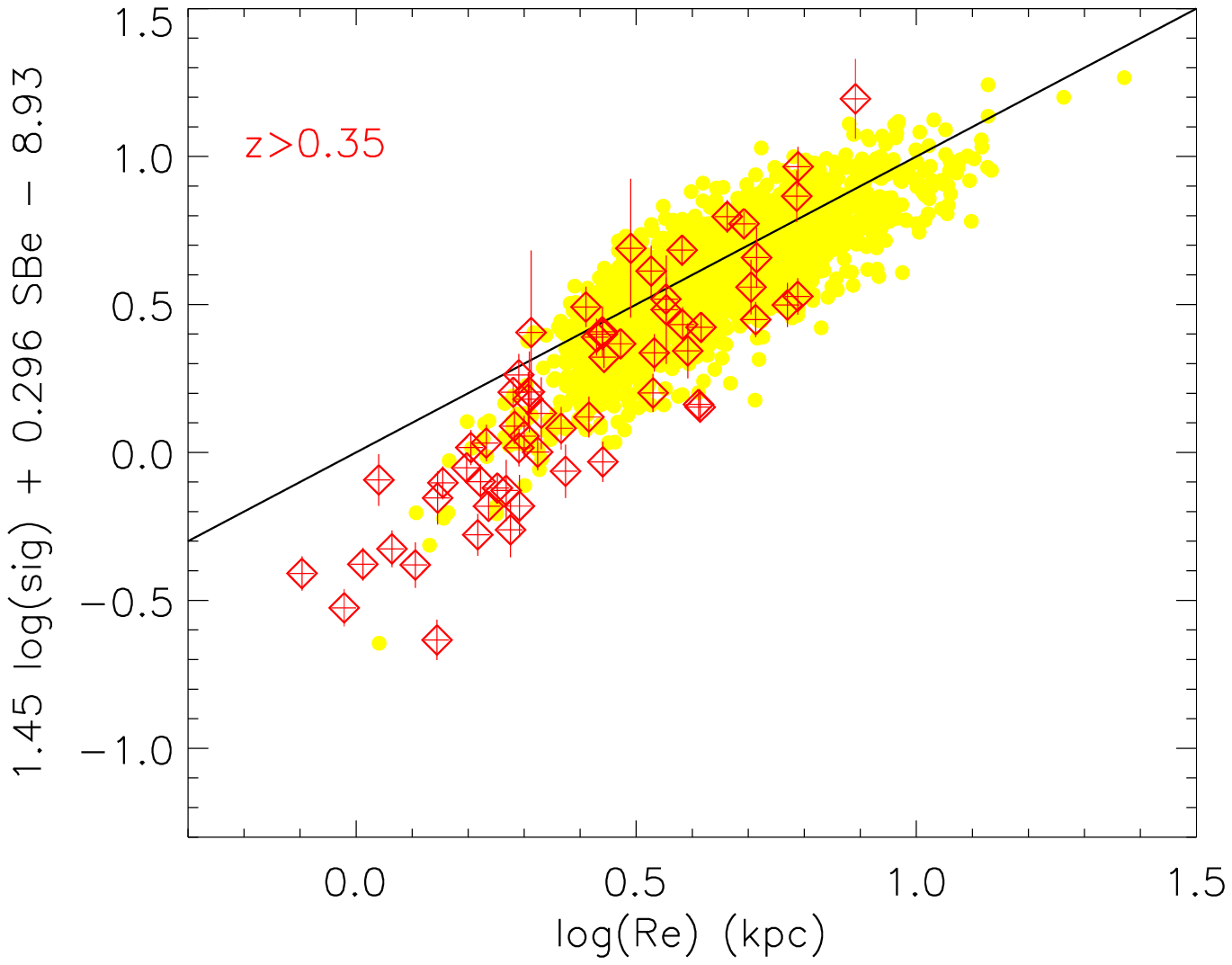}
      \caption{Edge--on projection of the Fundamental Plane in the g--band for the early--type galaxies in our sample. The black points represent the galaxies with z$<$0.35, and the open red diamonds are the objects with z$>$0.35. The small yellow points are the SDSS early--type galaxies of \citet{2003AJ....125.1866B}, represented using the FP coefficients obtained from the maximum likelihood ortogonal fit by these authors. In the left panel, we show the zero--point (ZP) and the rms scatter ($\sigma_{tot}$) for our local sample when the $a$ and $b$ coefficients are fixed to those of \citet{2003AJ....125.1866B}. In the right panel we present the same as for the left panel, but limiting our high--redshift sample and the SDSS one to the objects with absolute magnitudes in the range $-$21.5$>$M$_g$$>-$22.5.}
   \end{figure*}

To study the effects on the FP of these population of compact and bright objects, we used again the SDSS sample of \citet{2003AJ....125.1866B}. In the left panel of Fig.10, we present the FP in the g--band of our data compared to the SDSS maximum likelihood ortogonal fit. We found good agreement for our local sample of galaxies and for the high--redshift sample, the result is very similar to that was found for the B--band. In this case, from the change in the zeropoint of the FP, we found a brightening of 0.52 mag in the g--band for early--type galaxies at $<$z$>$=0.7. In the right panel of Fig.10 we represent the same FP but restricting our high--redshift sample and the SDSS one to the objects with absolute magnitudes in the range $-$21.5$>$M$_g$$>$$-$22.5. As in the rest of relations studied in the present work, the largest galaxies in this sample are distributed following the same FP as their local counterparts. Nevertheless, the smallest objects occupy an area of the plane that is poorly populated by local objects with the same characteristics. Note that the few objects of SDSS that have the same properties as our high--redshift compact objects, are distributed in a similar way in the FP, pointing to a decrease in this type of objects rather than other possible changes, such as luminosity differences caused by passive evolution. 

Although these compact and bright objects appear to be tilting the FP, our high--redshift sample is biased to the brightest objects. Therefore we cannot distinguish a change in the tilt from an increase in the scatter of the FP caused by these population of objects almost nonexistent in the local FP. 

\section{Summary and conclusions}

We have investigated the evolution of the FP in the B--band, using a large sample of 135 early--type galaxies selected from the EGS in the redshift range 0.2$<$z$<$1.2. Morphology was determined through visual classification using the V+I images of ACS. The structural parameters of these galaxies were obtained by fitting de Vaucouleurs stellar profiles to the ACS I--band images, using the {\tt GALFIT} code. To check the effect on the FP of the structural parameters obtained using different profiles, S\'ersic and bulge--to--disc decomposition models were also fitted to our sample of galaxies. We found good agreement in the FP derived from the three models, a result already found by \citet{2000ApJ...531..184K}. Finally, the velocity dispersion was calculated by extracting the stellar kinematics from the galaxy spectrum, using a maximum penalized likelihood approach.

Assuming that effective radii and velocity dispersions do not evolve with redshift, we found a different FP intercept for the high--redshift galaxies that could be interpreted as a brightening of 0.68 mag in the B--band and 0.52 mag in the g--band at $<$z$>$=0.7. However, the scatter in the FP for our high--redshift sample is substantially reduced when we allow the evolution of the FP slope, suggesting a different evolution of early--type galaxies according to their intrinsic properties, such as total mass, size or luminosity.

To investigate the galaxy properties responsible for the evolution of the FP, we derived the FJ and the Kormendy relations for our sample of galaxies. We obtained a preliminary result of galaxies being brighter in the past for a given velocity dispersion, and evidence of a stronger evolution for galaxies with low velocity dispersions. However, the great dispersion of this relation, does not allow deriving reliable conclusions. For the Kormendy relation, we found evolution in the surface brightness for a given effective radius, in the sense that galaxies were brighter in the past, with a change more noticeable for the smaller galaxies. Since this evolution could be caused by selection effects, we compared a subsample of our high--redshift objects with a subsample of the SDSS local galaxies of \citet{2003AJ....125.1866B} covering the same luminosity range. The surface brightness distribution as function of the effective radius of these objects is in agreement with those expected given their luminosity. However, we found a population of very compact (R$_e$$<$2 Kpc) and bright galaxies ($-$21.5$>$M$_g$$>$$-$22.5), of which only a small fraction (0.4$\%$) exist at z = 0, and that is responsible of the apparent evolution in the Kormendy relation. Then, a change in luminosity or size since z=1 is necessary to account for the decrease in number of this bright and compact objects.

For studying the processes driving the evolution of these objects, we analysed the luminosity--size and stellar mass--size relations. We found that the smallest objects present higher luminosities and stellar masses than their local counterparts, while the largest objects are similar to those at z=0. We ruled out passive evolution of stellar populations as responsible for this evolution since it cannot explain the change in the stellar mass. We think that the evolution of these compact objects is mainly caused by an increase in size, which could be explained by the action of $``$dry$"$ minor mergers that act increasing the size of the galaxies more than the mass.

Finally, we studied the effect of these bright and compact objects in the FP by comparing the galaxies with $-$21.5$>$M$_g$$>$$-$22.5 of our high--redshift sample and the SDSS one. The previous evolution found in the FP seems to be caused mainly by these galaxies, which have virtually disappeared at z=0. Unfortunately,
we cannot distinguish a change in the tilt from an increase in the scatter of the FP caused by these population of objects, since our high--redshift sample is biased to the brightest galaxies.

\begin{acknowledgements}

This work was supported by the Spanish {\em Plan Nacional de Astronom\'ia y Astrof\'isica} under grant AYA2008--06311--C02--01. We thank the DEEP2 group for making their catalogues and data publicly available. Funding for the DEEP2 survey has been provided by NSF grants AST95--09298, AST--0071048, AST--0071198, AST--0507428, and AST--0507483, as well as NASA LTSA grant NNG04GC89G. The work is based on observations obtained at the Canada--France--Hawaii Telescope (CFHT) which is operated by the National Research Council of Canada, the Institut National des Sciences de l'Univers of the Centre National de la Recherche Scientifique of France, and the University of Hawaii. This work is based in part on data products produced at the Canadian Astronomy Data Centre as part of the Canada--France--Hawaii Telescope Legacy Survey, a collaborative project of NRC and CNRS, and on observations obtained with MegaPrime/MegaCam, a joint project of CFHT and CEA/DAPNIA.

This study makes use of data from AEGIS, a multiwavelength sky survey conducted with the Chandra, GALEX, Hubble, Keck, CFHT, MMT, Subaru, Palomar, Spitzer, VLA, and other telescopes and supported in part by the NSF, NASA, and the STFC.

The Millenium Simulation database used in this paper and the web application providing online access to them were constructed as part of the activities of the German Astrophysical Virtual Observatory.

We thank the SAO/NASA Astrophysics Data System (ADS) that is always so useful.

 \end{acknowledgements}

\end{document}